\documentclass[reprint,floatfix,
showpacs,preprintnumbers,
amsmath,amssymb,aps,prd
]{revtex4-2}

\usepackage{multirow,hyperref}%
\usepackage[dvipsnames]{xcolor}%
\usepackage{multirow}%
\usepackage[final]{microtype}%
\usepackage{subcaption}%
\usepackage{amsmath,amssymb,amsfonts}%
\usepackage{amsthm}%
\usepackage{mathrsfs}%
\usepackage[title]{appendix}%
\usepackage{xcolor}%
\usepackage{textcomp}%
\usepackage{booktabs}%
\usepackage{algorithm}%
\usepackage{algorithmicx}%
\usepackage{algpseudocode}%
\usepackage{listings}%
\usepackage{graphicx}
\usepackage{dcolumn}
\usepackage{bm}

\def\be{\begin{equation}}
\def\ee{\end{equation}}
\def\beq{\begin{eqnarray}}
\def\eeq{\end{eqnarray}}

\def\n{\widetilde{\nabla}}

\def\T{\bar{\Theta}}

\def\n{\nonumber}

\def\R{\mathcal{R}}
\def\TK{T_{\text{\tiny K}}}
\def\rd{\rho_{\text{\tiny dark}}}
\def\re{\rho_{\text{\tiny eff}}}
\def\Odr{\Omega_{dr,0}}
\def\Ob{\Omega_{\beta,0}}
\def\Om{\Omega_{m,0}}
\def\Ok{\Omega_{k,0}}
\def\Pe{P_{\text{\tiny eff}}}
\def\ome{\omega_{\text{\tiny eff}}}
\def\Oall{\Omega_{m,\beta,dr,z}}
\def\oall{\omega_{m,\beta,z}}
\def\Pall{\Phi_{m,\beta,dr,z}}

\begin{document}




\title{Dark energy as a geometrical
effect in geodetic brane gravity}


\author{Efra\'\i n Rojas}
\email{efrojas@uv.mx}

\author{G. Cruz}
\email{giocruz@uv.mx}

\author{J. C. Natividad}
\email{julio19natizaca@gmail.com}

\affiliation{Facultad de F\'\i sica, Universidad Veracruzana, 
Paseo No. 112, Desarrollo Habitacional Nuevo Xalapa, Xalapa-Enr\'\i quez, 91097, Veracruz, M\'exico
}%






\begin{abstract}
Within the framework of the modified geodetic brane gravity, 
conformed by the Regge-Teitelboim model and enhanced with 
a linear term in the extrinsic curvature of the brane,  
the possibility that under an FRW geometry this theory 
emulates the so-called dark energy is discussed. The cosmological
behavior of this model displays a self-(non-self)-accelerated 
expansion of this universe which is caused by a combination 
of usual matter and gravitational geometric effects controlled 
by a $\beta$ parameter that accompanies the correction $K$ 
term. Indeed, the self-accelerated branch, provided by the trace 
$K$ model raises the question of whether the 
extrinsic curvature correction terms might be suitable for dark 
energy candidates. We discuss the analytical expression obtained 
for $\rd$ in addition to the main cosmological parameters such 
as the state parameter $\omega_{\text{\tiny eff}}$ and the 
deceleration parameter $q$. Moreover, when we call for the 
contribution of dark radiation-like energy to be switched off, 
$\Odr \to 0$, we  find the same acceleration behavior, as 
well as the same dark energy content provided by the DGP 
theory. 
The relationship of our findings to the analysis for $\rd$ performed by Davidson and Gurwich within the unified brane cosmology is briefly discussed.    
 
\end{abstract}


\maketitle


\section{\label{sec:intro}Introduction}

Any attempt to solve the still unexplained mystery of the 
dark matter/energy,  as long as it is based on fundamental 
geometric symmetries which play a central role in respected 
gravitational theories, is welcome. The dark energy nature 
constitutes a tantalizing question to which the cosmology 
and the particle physics are attempting to provide a 
compelling answer. Following this line of reasoning, within 
the geodetic brane gravity based in the Regge-Teitelboim 
(RT) model, we would like to explore a possible explanation 
to such a puzzle where extrinsic curvature terms can provide 
findings that reproduce many of the characteristics that 
exhibits the so-called dark matter/energy. 
The RT gravity theory generalizes General Relativity (GR) in 
a particular theoretical direction~\cite{RT1975}. The theory 
is based on a simple particle/string-inspired assumption that 
our Universe is a 4-dimensional surface isometrically 
embedded, and evolving, in a higher-dimensional flat ambient 
spacetime. In this view, our 4-dimensional spacetime is 
regarded as the trajectory (worldvolume) of a 3-dimensional 
extended object, probing a higher-dimensional flat ambient 
space. Within this assumption, one must not overlook the local 
isometric embedding theorem~\cite{Friedman1961,Rosen1965, Kasner1921}; indeed, in order to ensure the local 
existence of an embedding framework, at most $N = n(n +1)/2$ background flat dimensions are required to locally embed a $n$-metric. Clearly, for a $4$-dimensional brane, at most a $10$-dimensional flat ambient spacetime is 
needed, a number that can be reduced when the brane metric 
admits some Killing vector fields as fortunately occurs in 
the case of an FRW brane geometry. In this sense, as one learns for example in~\cite{Rosen1965}, with $N=5$ 
background flat dimensions it is sufficient. The RT action is identical to the Einstein-Hilbert (EH) action 
of GR but, with the crucial difference that the embedding 
functions $X^\mu(x^a)$ rather than the metric $g_{ab}$ 
defined of the worldvolume, are elevated to the level of field 
variables. The field equations result in the Einstein 
ones but contracted with the extrinsic curvature tensor 
of the embedding which represents a generalization of 
the Einstein equations in the sense that the RT equations 
are fulfilled for every solution of the Einstein equations. 
The theory has attractive properties such as the already
mentioned built-in Einstein limit and in particular, within 
the braneworld context, a dark companion to any primitive 
energy density. Under the names of \textit{geodetic 
brane gravity} or \textit{embedding gravity}, the RT model 
has been widely investigated by many authors as an interesting 
formulation to describe gravity, cosmological implications 
in extra dimensions as well as a suitable alternative 
for quantization of gravity~\cite{Deser1976,Tapia1989,Maia1989,Davidson1998,
Davidson2003,Pavsic2002,Paston2007,Rojas2009,Estabrook2010,Sheykin2023,
Biswajit2014,Rojas2022,Stern2022,Paston2023,Stern2023}. 

The trace $K$ model, a linear term in the extrinsic 
curvature, represents an alternative term such as RT 
gravity based on the isometrical embeddings that might 
underlie cosmic acceleration, and which does not alter 
the order of the equation of motion~\cite{Rojas2012}. 
In this spirit, the dynamics of the brane is still 
driven by the Gauss-Codazzi, Codazzi-Mainardi, and Ricci 
integrability conditions of the geometry of surfaces~\cite{Spivak1979,defo1995}.
Certainly, the trace $K$ term offers some singular 
features that deserve full interest. For instance, within 
the braneworld cosmological scenario, this model acts 
like a fluid, emulating effects such as that of a 
cosmological constant, and a self-(non-self)-acceleration 
dynamical behavior at late times and producing unnatural 
matter. 

In this paper, within the extended objects framework, we 
attempt to go one step beyond pure RT gravity and reignite 
the discussion regarding the dark energy companion of the 
primitive $\rho$ for the RT cosmology as discussed in~\cite{Davidson1999,Davidson2001,Davidson2005}, by 
exercising the geometrical features contained in the trace 
$K$ model. Specifically, in this regard, we show that a 
kind of dark energy is a geometrical effect. 
Certainly, some braneworld models at low energies have 
become interesting scenarios providing late-time acceleration 
behaviors without the need to consider dark energy by 
adding in the extra mysterious matter. We observe that 
$\beta$, the parameter accompanying the trace $K$ model, 
is the only input parameter in our development that produces 
this effect. In this sense, with the claim of exploring the 
source of the dark energy, within the modified gravity theories 
context, one of the requirements is the built-in Einstein limit. On the basis of geodetic brane gravity, it is interesting and 
motivating to improve the theory by incorporating properly 
pathologies-free geometric terms so that physical effects 
from other more elaborated theories can be reproduced. A pioneer work in this direction has been performed in~\cite{Davidson2006} within the framework of unified brane cosmology where Davidson and Co., find a 
dark energy expression representing a dark companion of 
the observable $\rho$. Despite 
the bulk being flat Minkowski, there are local effects on the 
brane via conserved quantities associated with the Poincar\'e 
invariance, and the correction $K$ term which can be thought 
of as a remnant of the existence of a 5-dimensional EH action. 
We find that our proposal, when applied to a homogeneous and 
isotropic universe that has an FRW geometry, leads to a suggestive 
modification of the Friedmann equation which includes an 
additional term, $\rd$, in the effective energy density and 
that it is treated in the manner that the primitive $\rho$.
The presence of such $\rd$ originates mainly from the 
extrinsic curvature correction term $K$ and the matter 
confined to the braneworld. The trace $K$ model requires 
careful observation; needles to say such a term resembles 
the Gibbons-Hawking-York (GHY) counterterm but with the 
essential difference that in our approach $K$ contains 
time derivatives of the field variables that contribute 
to the dynamics of the brane. It is worth mentioning this 
model has cropped up in other contexts ranging from lipid 
membranes to particle physics and brane-world gravity~\cite{Svetina1989,Onder1988,Rojas2011}.

In the first part of our work, our strategy is, under 
an FRW geometry on the brane, to make use of the invariance 
under reparametrizations of the model and identify 
conserved quantities that help to integrate the equation 
of motion, and thus lead us to identify a Friedmann type 
equation. By invoking, and imposing, the traditional point 
of view of the cosmology for the Friedmann equation, we 
may identify an effective energy density where $\rd$ is 
a companion piece of the primitive energy density $\rho$. 
Indeed, we can express the dark energy contribution in 
terms of the associated extrinsic curvature correction 
parameter, $\beta$, and the confined matter, producing a 
self-(non-self)-acceleration behavior of this type of 
universe. One could wonder if the answer to the 
tantalizing question of dark energy is related to the 
inherent extrinsic curvature and its embedding properties, 
as was perceived in~\cite{Maia2005}, so our approach 
suggests that is the answer. On the other 
hand, the intricate analytical expression that we found to 
know $\rd$, Eq.~(\ref{eom4}), coincides with that previously 
obtained by Davidson and Gurwich in~\cite{Davidson2006} making use 
of a Dirac's brane variation prescription within the unified 
brane theory framework.
 Subsequently, we analytically 
compute important cosmological parameters such as the 
equation of state parameter and the deceleration parameter. 
These confirm the existence of a type of dark energy 
density and the existence of a self-(non-self)-acceleration 
behavior through the emergence of Big Chill and Big Bounce 
histories of this universe. It is worth mentioning that 
this model suffers from a fine-tuning problem, so one must 
rely on numerical analysis.

This paper is organized as follows. In Sec.~\ref{sec2} 
we review the geometrical aspects of the geodetic brane 
gravity enhanced with the trace $K$ model. We discuss the 
resulting equation of motion, the built-in Einstein limit, 
and the definition of the concept of embedding matter. 
Additionally, we accommodate our development to an FRW 
geometry. In Sec.~\ref{sec3} we obtained an analytical 
expression for the dark energy contribution in our framework 
as well as some cosmological parameters of interest. Further, 
we specialize our analysis to the DGP approach when the 
contribution from the extra dimension is turned off. 
Finally, in Sec.~\ref{sec4} we conclude and discuss our 
findings. 

\section{Modified Regge-Teitelboim gravity}
\label{sec2}

The modified Regge-Teitelboim gravity,~\cite{Rojas2012}, 
including a matter content, is provided by
\be 
S[X^\mu] = \int_m d^{3+1} x\,\sqrt{-g}\,\left( 
\frac{\alpha}{2} \mathcal{R} + \beta\,K 
+ L_{\text{\tiny matt}}\right),
\label{action}
\ee
where $\alpha$ and $\beta$ are constants with appropriate 
dimensions. The scalars $\R$ and $K$ are formed using the
induced metric $g_{ab}$ and the extrinsic curvature $K_{ab}$,
that encode the geometrically most important derivatives of
the embedding functions $X^\mu$,
\beq 
g_{ab} &=& \eta_{\mu\nu} \partial_a X^\mu \partial_b X^\nu,
\label{gab}
\\
K_{ab} &=& - \eta_{\mu\nu} n^\mu \nabla_a \partial_b X^\nu.
\label{Kab}
\eeq
Here, $\partial_a X^\mu =: e^\mu{}_a$ and $n^\mu$ represent the 
tangent vectors and the normal vector to the world volume $m$, respectively, and $\nabla_a$ is the covariant derivative compatible with 
$g_{ab}$, and $\eta_{\mu\nu}$ is the flat Minkowski metric
in the background spacetime. $\mu = 0,1, \ldots, 4$ 
and $a=0,1,2,3$ (see Appendix~\ref{app1} for details about 
of the geometrical constructions behind the description of 
extended objects). Certainly, $\R$ is the Ricci scalar defined 
on $m$ (our Universe) obtained from $g_{ab}$, and 
$K=g^{ab} K_{ab}$ denotes the trace of $K_{ab}$ where $g^{ab}$ 
stands for the inverse of $g_{ab}$. Further, $g = \det 
(g_{ab})$. The possible matter fields coupled to $m$ are 
embodied in the Lagrangian $L_{\text{\tiny matt}}$. 

The action~(\ref{action}) resembles Einstein-Hilbert action 
defined on $m$, augmented with the dynamical term $K$,
but with the difference that the independent variables are the 
embedding functions $X^\mu (x^a)$, where $x^a$ are local 
coordinates of $m$. In this sense, the RT gravity is, by 
construction, a continuation of string theory so it justified 
that the field variables are the $X^\mu$ and not the induced 
metric components $g_{ab}$.

The variational principle applied to $S$, under infinitesimal 
changes $X^\mu (x^a) \rightarrow X^\mu (x^a) + \delta X^\mu 
(x^a)$, leads to the fact that the physical deformations of 
the action are the normal to $m$,  $\delta_\perp S = 
0$,~\citep{defo1995,Capo2000,Rojas2013} so 
that the only field equation in a coordinate-invariant form is
\be 
\left( \alpha G^{ab} + \beta \TK^{ab} - T^{ab}_{\text{\tiny 
m}}\right) K_{ab} = 0,
\label{eom1}
\ee
where $G_{ab} = \R_{ab} - \frac{1}{2} g_{ab} \R$ is the
worldvolume Einstein tensor, $T_{ab}^{\text{\tiny K}} := 
K_{ab} - g_{ab}K$, and $T_{ab}^{\text{\tiny m}} := 
\frac{(-2)}{\sqrt{-g}} \frac{\partial (\sqrt{-g}\,
L_{\text{\tiny matt}})}{\partial g^{ab}}$ is the worldvolume 
energy-momentum tensor. Although the action~(\ref{action}) is 
of second order in derivatives of $X^\mu$, equation~(\ref{eom1}) 
represents a second-order in derivatives equation of motion. 
In passing, by using the contracted Codazzi-Mainardi 
equation,~(\ref{CMequation}), one verify that $\nabla_a 
\TK^{ab} = 0$, in analogy with the usual 
Bianchi identity $\nabla_a G^{ab} = 0$.

With support on the invariance under reparametrizations 
of $m$, that is the symmetry of the action~(\ref{action}), 
the mechanical content of $m$ can also be obtained 
from the current conservation law, $\nabla_a f^{a\,\mu} = 0$, 
\cite{Capo2000}, with
\be 
f^{a\,\mu} = -  \left( \alpha G^{ab} + \beta \TK^{ab} 
- T^{ab}_{\text{\tiny m}}\right) e^\mu{}_b,
\label{famu}
\ee
being the conserved stress tensor which is purely tangential
to the world volume. Indeed, $f^{a\,\mu}$ obeys
\beq 
& - &\nabla_a \left( \alpha G^{ab} + \beta \TK^{ab} - 
T^{ab}_{\text{\tiny m}}\right) e^\mu{}_b 
\n 
\\
&+&
\left( \alpha G^{ab} + \beta \TK^{ab} - 
T^{ab}_{\text{\tiny m}}\right) K_{ab} \,n^\mu = 0.
\n
\eeq
From this, we immediately infer the equation of 
motion~(\ref{eom1}) as well as the fact that 
$T^{ab}_{\text{\tiny m}}$ is covariantly conserved, 
$\nabla_a  T^{ab}_{\text{\tiny m}} = 0$. 

Some comments are in order. The scalar $K$ that appears in 
the action~(\ref{action}) is similar to the GHY counterterm 
in pure GR but this analogy needs a word of caution. Within 
the framework of extended objects, floating in an ambient
spacetime, the $K$ term possesses time derivatives of the 
variables $X^\mu$, see~(\ref{Kab}); hence, such a term provides 
dynamics on the hypersurface $m$. Regarding this, $\TK^{ab}$ is
a purely geometric tensor that plays the role of a 
energy-momentum-like tensor. In this line of reasoning
the $K$ term can be thought of as a matter type contribution 
to the Einstein-Hilbert term as we will see below. On 
the other hand, in view of~(\ref{eom1}), we introduce
\be 
\mathsf{T}^{ab}:= \alpha G^{ab} + \beta \TK^{ab} -  
T^{ab}_{\text{\tiny m}}.    
\label{Ttensor}
\ee 
This entails that the equation of motion~(\ref{eom1}) can be 
written in the form
\be 
\left[ G^{ab} - \frac{1}{\alpha} 
\left( 
T^{ab}_{\text{\tiny m}} - \beta \TK^{ab}
\right) 
\right]K_{ab} = 0, 
\ee 
or,  
\be
\nabla_a \left( \mathsf{T}^{ab}\,\partial_b X^\mu 
\right) = 0.
\label{eom2}
\ee
Therefore, these imply that, on the one hand,
if $\mathsf{T}^{ab} = 0$ with $\beta = 0$, every solution 
of Einstein equations is automatically a solution of 
the Regge-Teitelboim gravity; on the other hand, 
if $\mathsf{T}^{ab} \neq 0$, different solutions are 
allowed to this type of gravity. Regarding this last point,
as discussed in~\cite{Davidson2001,Paston2018}, $\mathsf{T}^{ab}$
must be understood as an additional fictional matter 
contrasting with the ordinary one, and labelled as 
\textit{dark matter} or \textit{embedding matter}, making 
it a possible explanation for dark matter. 
In connection with this, notice that 
\be 
\nabla_a (T^{ab}_K\,\partial_b X^\mu) = 0,
\qquad \text{or} \qquad 
\partial_a \left( \sqrt{-g}\,j^{\mu\,a}_K \right)
= 0,
\ee
with the geometric current $j^{\mu\,a}_K$ defined by
\be 
j^{\mu\,a}_K := \left( K^{ab} - K \,g^{ab} \right) 
\partial_b X^\mu.
\ee

\subsection{Friedmann cosmology in modified RT gravity}
\label{sub:1}

We will assume that our universe is the evolution of a 
$3$-dimensional extended object, say $\Sigma$, in a 
$5$-dimensional Minkowski spacetime
parametrized in spherical coordinates, 
$ds^2_5 = - dt^2 + da^2 + a^2 d\Omega^2_3$, where 
$d\Omega^2_3 = d\chi^2 + \sin^2 \chi d\theta^2 
+ \sin^2\chi \sin^2 \theta d\phi^2$ 
denotes the unit $3$-sphere. At this stage a parametric 
representation for this universe $m$, assuming homogeneity, 
isotropy and a closed topology is provided by
\be 
\label{embedding}
x^\mu = X^\mu (x^a) = (t (\tau), a (\tau), \chi, \theta, \phi),
\ee
where $x^\mu$ are the local coordinates for the ambient spacetime
and $\tau$ denotes the proper time for an observer at rest 
concerning $m$. Recall, $a,b = 0,1,2,3$.
The orthonormal basis adapted to $m$ is provided by the four
tangent vectors $e^\mu{}_a = \partial X^\mu / \partial x^a$ 
complemented with the unit spacelike normal vector $n^\mu = 
(1/N) (\dot{a}, \dot{t},0,0,0))$ where $N := \sqrt{\dot{t}^2 
- \dot{a}^2}$. Here and henceforth, an overdot denotes a 
$\tau$-derivative. The metric induced on $m$ from the 
ambient spacetime,~(\ref{gab}), is
\be 
ds^2_4 = g_{ab} dx^a dx^b = - N^2 d\tau^2 + a^2 d\Omega^2_3,
\label{metric}
\ee
while the non-trivial components of the extrinsic 
curvature~(\ref{Kab}) are
\be 
\label{Kab1}
K^\tau{}_\tau = \frac{\dot{t}^2}{N^3} \frac{d}{d\tau}
\left( \frac{\dot{a}}{\dot{t}}\right),
\quad \quad
K^\chi{}_\chi = K^\theta{}_\theta = K^\phi{}_\phi
= \frac{\dot{t}}{\dot{Na}}.
\ee
The Gauss-Codazzi equations~(\ref{GCequation}), allows 
us to know the world volume Riemann and Ricci tensors, 
and the Ricci scalar, in terms of the extrinsic curvatures. 
Hence,
\beq 
K &=& \frac{1}{N^3} (\dot{t} \ddot{a} - \dot{a}\ddot{t})
+ \frac{3\dot{t}}{Na},
\label{K}
\\
\R &=& \frac{6 \dot{t}}{N^4 a} (\dot{t} \ddot{a} - \dot{a}
\ddot{t}) + \frac{6\dot{t}^2}{N^2 a^2}.
\label{R}
\eeq 
It is required that $T^{ab}_{\text{\tiny m}}$ acquire 
the form of a general perfect fluid
\be 
T^{ab}_{\text{\tiny m}} = \left( \rho + P \right)
\eta^a \eta^b + P\, g^{ab},
\label{Tab1}
\ee 
with energy density $\rho (a)$ and pressure $P(a)$, and 
$\eta^a$ being the timelike unit normal vector to the 
extended object at a fixed $\tau$. 

The non-vanishing components of the conserved tensors 
$T^{ab}_{\text{\tiny m}}$, $\TK^{ab}$ and $G_{ab}$
can be computed for the Friedmann-like geometry in a
straightforward manner
\begin{widetext}
\be 
\begin{aligned}
T_{\text{\tiny m}}^\tau{}_\tau &= - \rho
& \qquad T_{\text{\tiny m}}^\chi{}_\chi &= 
T_{\text{\tiny m}}^\theta{}_\theta
= T_{\text{\tiny m}}^\phi{}_\phi = P,
\\
\TK^\tau{}_\tau &= - \frac{3\dot{t}}{Na}
 & \qquad \TK^\chi{}_\chi &= \TK^\theta{}_\theta
 = \TK^\phi{}_\phi = - \frac{\dot{t}^2}{N^3}
 \frac{d}{d\tau} \left( \frac{\dot{a}}{\dot{t}}\right)
 - \frac{2\dot{t}}{Na},
\\
G^\tau{}_\tau &= - \frac{3\dot{t}^2}{N^2a^2}
 & \qquad G^\chi{}_\chi &= G^\theta{}_\theta
 = G^\phi{}_\phi = - \frac{2\dot{t}^3}{N^4 a}
 \frac{d}{d\tau} \left( \frac{\dot{a}}{\dot{t}}\right)
 - \frac{\dot{t}^2}{N^2 a^2}.
\end{aligned}
\label{TandG}
\ee
\end{widetext}
As discussed in~\cite{Rojas2012}, for example, the flat 
and hyperbolic geometries can be included by convenient 
generalizations of the embedding~(\ref{embedding}). To 
accommodate these geometries in the cosmological development 
for this model, we just need to do the promotion $\dot{t} 
\longrightarrow (\dot{a}^2 + k)^{1/2}$ with $k = -1,0,1$, 
accompanied by the usual cosmic gauge with a function 
$N \longrightarrow \mathcal{N}$, specific to each embedding.

By suitably writing the equation of motion~(\ref{eom3}),
we wish to address the analysis of the mentioned embedding 
matter equation of motion~(\ref{eom2}) when 
$a=\tau$ and $\mu=t$ and obtain
\be 
\frac{1}{\sqrt{-g}} \partial_\tau \left( \sqrt{-g}\,
\mathsf{T}^{\tau\tau}\,\partial_\tau X^t \right) = 0.
\label{eom3}
\ee
This is accompanied by the energy-momentum conservation 
law $\nabla_a T^{ab}_{\text{m}} = 0$ which produces the 
integrability condition $\dot{\rho} + 3\left( 
\frac{\dot{a}}{a}\right)(\rho + P) = 0$. In view of this, from 
the geometric conservation law $\nabla_a T^{ab}_K = 0$, 
one could expect a cosmological energy-momentum conservation 
equation related to the trace $K$ model. We wish to address the 
solving analysis of these equations to explore the dark 
matter content provided by this brane model, as we will see shortly.

\section{Dark energy from modified geodetic brane cosmology}
\label{sec3}

The solution provided by~(\ref{eom3}) leads to a constant
of motion. To obtain this we first insert the 
values~(\ref{TandG}) into the definition~(\ref{Ttensor}) to get
$\mathsf{T}^{\tau\tau} = \frac{3\alpha \dot{t}^2}{N^4 a^2}
+ \frac{3\beta \dot{t}}{N^3 a} - \frac{\rho(a)}{N^2}$. Similarly,
by considering $g = -N^2 a^6 \sin^4 \chi \sin^2 \theta$ and 
then inserting these results into expression~(\ref{eom3}) we get
\be 
\n
\partial_\tau \left[ \frac{3\alpha\, a \,\dot{t}^3}{N^3}
+ \frac{3\beta\, a^2 \dot{t}^2}{N^2} - \frac{a^3 \,\rho 
(a)\,\dot{t}}{N}
\right] = 0.
\ee
After a trivial integration followed by choosing the cosmic 
gauge, $N=1$, and the inclusion of the three main geometries 
by $\dot{t} \longrightarrow (\dot{a}^2 + k)^{1/2}$, we arrive
at our master equation
\be 
\label{eom4}
-3\alpha a\,(\dot{a}^2 + k)^{3/2} - 3\beta a^2\,(\dot{a}^2
+ k) + \rho a^3\,(\dot{a}^2 + k)^{1/2} =: 
\frac{\omega}{\sqrt{3\,\alpha}},
\ee
with $\omega$ being a constant. In fact, $\omega$ plays a 
double duty. On the technical side, it is related to the conserved 
bulk energy conjugate to the embedding time coordinate 
$t(\tau)$ and, on the physical side, within the unified brane cosmology~\cite{Davidson2006}, it parameterizes the 
deviation from the Randall-Sundrum brane cosmology and from 
general relativity when $\omega = 0$ in addition to 
$\beta = 0$. With an eye to cosmology, this last remark 
represents the main role of the $\omega$ constant.
An additional remark is in order; as discussed 
in~\cite{Davidson2006} $\omega$ is not necessarily a small 
quantity and its size will be fixed once the dark matter 
interpretation is fully established.
Further, after some algebra and solving 
an involved cubic equation,~(\ref{eom4}) leads to an effective 
Friedmann equation~\cite{Rojas2012}.

If we do not renounce the point of view of the standard cosmology,
ignoring the existence of an extra dimension and the brane 
gravity, we would rearrange~(\ref{eom4}) so that it can be cast 
in the form
\be 
\dot{a}^2 + k = \frac{1}{3\alpha} \left( \rho + \rd
\right) a^2,
\label{FRW1}
\ee
where $\rd$ encode all the extra contributions, physical and 
geometrical, provided by the extra dimension to the primitive 
energy density $\rho$. 
In this sense, the evolution of this type of universe
is dictated by an effective energy density $\re: = \rho 
+ \rd$ and not solely by the primitive energy density $\rho$.
Further, the Friedmann-like equation~(\ref{FRW1}) can be
written in the fashion $\dot{a}^2 + k = \frac{1}{3\alpha}
f(\rho)a^2$ once we determine the form of $\rd$ as we will see
shortly. It is worth mentioning that the form~(\ref{FRW1}) 
can also be obtained by assuming $T^{ab}_K$ has a contribution 
similar to that of a perfect fluid of the form $T_K^\tau{}_\tau = 
\rho_K$ and $T_K^i{}_i = P_K$, with $i= \chi,\theta,\phi$ 
along with the requirement $\dot{\rho}_K + 3\frac{\dot{a}}{a} 
(\rho_K + P_K) = 0$, \cite{Stern2022}.

If~(\ref{eom4}) and~(\ref{FRW1}) are in accordance, 
$\rd$ 
results in the root of a cubic equation. 
Certainly, plugging~(\ref{FRW1}) into~(\ref{eom4}) yields
\be 
\n
3\beta_*\,(\rho + \rd) + \frac{\omega}{a^4} = - \rd \left( 
\rho + \rd \right)^{1/2},
\ee
where we have put $\beta_* := \frac{\beta}{\sqrt{3\,\alpha}}$. 
By defining $\mathcal{Z} := (\rho + \rd)^{1/2}$ and substituting 
it into the latter equation, followed by a reorganization, we find
\be 
\label{rdcubic1}
\mathcal{Z}^3 + 3\,\beta_* \,\mathcal{Z}^2 - \rho\,\mathcal{Z} + 
\frac{\omega}{a^4} = 0.
\ee
This is precisely the implicit equation for the dark matter
content that we intend to analyze.  In passing, we see that
for $\beta_* = 0$ we get $\rd^2 (\rho + \rd) = 
\frac{\omega^2}{a^8}$ in complete agreement with the result 
found in~\citep{Davidson1999,Davidson2001,Davidson2005}. 
By invoking the change of variable $\varrho := \mathcal{Z} 
+ \beta_*$ in~(\ref{rdcubic1}) we find the depressed 
cubic equation $\varrho^3 + p(a)\,\varrho + q(a) = 0$ 
where
\be 
p = - \left( \rho + 3 \,\beta_*^2 \right),
\qquad \text{and} \qquad
q =  \beta_* \left( \rho + 2 \,\beta_*^2 \right)
+ \frac{\omega}{a^4}.
\label{p}
\ee
The solution for the depressed cubic equation can be written 
in terms of trigonometric or hyperbolic functions with the use 
of a couple handy identities, $4 \cos^3 \theta - 3 \cos \theta 
= \cos (3 \theta)$ or $4 \cosh^3 \theta - 3 \cosh \theta = \cosh 
(3\theta)$. Certainly, depending on the values chosen for 
functions~(\ref{p}) we could have solutions with physical interest.
In any case, the general forms of the solutions are
provided by
\be 
\rd = \left[ 2 \sqrt{\frac{|p|}{3}} \cosh \left( 
\frac{1}{3} \cosh^{-1} (\phi) \right) - \beta_* 
\right]^2 - \rho,
\label{sol1}
\ee
or 
\be 
\rd = \left[ 
2 \sqrt{\frac{|p|}{3}} \cos \left( \frac{1}{3} \cos^{-1}
(\phi) + n \frac{2\pi}{3}\right) - \beta_*
\right]^2 - \rho,
\label{sol2}
\ee
with $n=0,1,2$. Here, $\phi := - \frac{q}{2} \left( - 
\frac{3}{p} \right)^{3/2}$.  
More precisely,~(\ref{sol1}), applies when $p \leq 0$ and 
$|\phi| \geq 1$, while~(\ref{sol2}), will be the one of 
interest when $p \leq 0$ and $|\phi| \leq 1$. Additionally, 
if $p > 0$ we would have another possible scenario; in such a 
case the main function will be $\sinh$ instead of $\cosh$ or $\cos$. 
We confine ourselves to the physical solutions
provided by~(\ref{sol1}) or~(\ref{sol2}). Note that whatever 
the physical solution is, the definite positivity of the 
total energy density holds, which is one of the exclusive features 
of the geodetic brane gravity~\cite{Davidson2005}. Notice in passing 
that this conclusion holds even for $\rd < 0$. 
Additionally, since $T^{ab}_K$ is covariantly conserved,
and insisting on the cosmological principle, we assume that
$P_{\text{\tiny dark}}$ can be obtained from~(\ref{sol1})
or~(\ref{sol2})
\be 
P_{\text{\tiny dark}} = - \left( \frac{a}{3\dot{a}}\,
\dot{\rho}_{\text{\tiny dark}} + \rd \right).
\label{Pd}
\ee

We find convenient to rewrite expressions~(\ref{sol1}) 
and~(\ref{sol2}) in terms of energy density parameters. 
Indeed, for simplicity, from now on we restrict ourselves 
to consider dust matter content on the brane so that 
$\rho(a) := \frac{\rho_0}{a^3}$, with $\rho_0$ being a 
constant. When introducing the Hubble parameter 
$H:= \frac{\dot{a}}{a}$ as well the energy density
parameters 
\be 
\label{denergies}
\Om := \frac{\rho_0}{3\alpha H_0^2},
\quad \Ob := \frac{\beta}{\alpha\,H_0},
\quad \Odr := \frac{\omega}{(3\alpha)^{3/2} H_0^3},
\ee
where as usual, $H_0$ is the Hubble constant,~(\ref{sol1}) 
and~(\ref{sol2}) turn into the expressions
\begin{widetext}
\be
\frac{\rd}{\alpha H_0^2} = - \frac{3\Om}{a^3}
+ \frac{1}{3} \left\lbrace
2\sqrt{\Ob^2 + \frac{3\Om}{a^3}} 
\cosh \left[
\frac{1}{3} \cosh^{-1} \left( - \frac{\Ob \left( \Ob^2 + \frac{9}{2} 
\frac{\Om}{a^3}\right) + \frac{27}{2} \frac{\Odr}{a^4}}{\left( 
\Ob^2 + \frac{3 \Om}{a^3}\right)^{3/2}} \right)
\right] - \Ob
\right\rbrace^2,
\label{sol3}
\ee
and
\be
\frac{\rd}{\alpha H_0^2} = - \frac{3 \Om}{a^3}
+ \frac{1}{3} \left\lbrace
2\sqrt{\Ob^2 + \frac{3 \Om}{a^3}}  
\cos \left[
\frac{1}{3} \cos^{-1} \left( - \frac{\Ob \left( \Ob^2 + \frac{9}{2} 
\frac{\Om}{a^3}\right) + \frac{27}{2} \frac{\Odr}{a^4}}{\left( 
\Ob^2 + \frac{3 \Om}{a^3}\right)^{3/2}} \right)
+ n \frac{2\pi}{3} \right] - \Ob
\right\rbrace^2,
\label{sol4}
\ee
\end{widetext}
where $n=0,1,2$.
The choice of one of the solutions is in dependence of the values
of the density energy parameters~(\ref{denergies}) satisfying the 
inequalities provided by the discriminant of the cubic equation.
Indeed, for the depressed cubic equation, the discriminant is given
by $\Delta = - \left( 4 p^3 + 27 q^2 \right)$. If $\Delta < 0$
we shall have one real root provided by~(\ref{sol3}) and two 
non-real complex conjugate roots while that for $\Delta > 0$ we 
shall have three distinct real roots provided by~(\ref{sol4}).
Considering~(\ref{p}) and~(\ref{denergies}), it is straightforward 
to show that for
\begin{widetext}
\be
\frac{\Ob \left( \Ob^2 + \frac{9}{2} \frac{\Om}{a^3}\right) 
+ \frac{27}{2} \frac{\Odr}{a^4}}{\left( \Ob^2 
+ \frac{3 \Om}{a^3}\right)^{3/2}}  > 1, 
\qquad \quad \text{or} \qquad \quad 
\frac{\Ob \left( \Ob^2 + \frac{9}{2} 
\frac{\Om}{a^3}\right) + \frac{27}{2} \frac{\Odr}{a^4}}{\left( 
\Ob^2 + \frac{3 \Om}{a^3}\right)^{3/2}} < 1,
\label{conditions}
\ee 
\end{widetext}
we have one physical root or three physical ones, respectively.

\begin{figure*}[!hbtp]
\begin{subfigure}{0.325\linewidth}
    \centering
    \includegraphics[width=\linewidth]{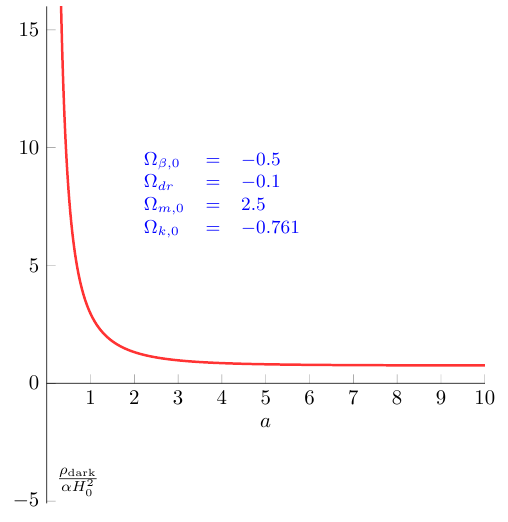}
    \caption{}
    \label{fig:y1}
\end{subfigure}
\hfill
\begin{subfigure}{0.325\linewidth}
    \centering
    \includegraphics[width=\linewidth]{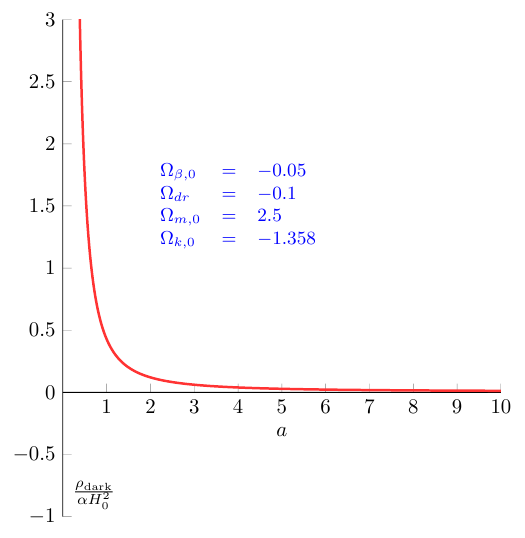}
    \caption{}
    \label{fig:y2}
\end{subfigure}
\hfill
\begin{subfigure}{0.325\linewidth}
    \centering
    \includegraphics[width=\linewidth]{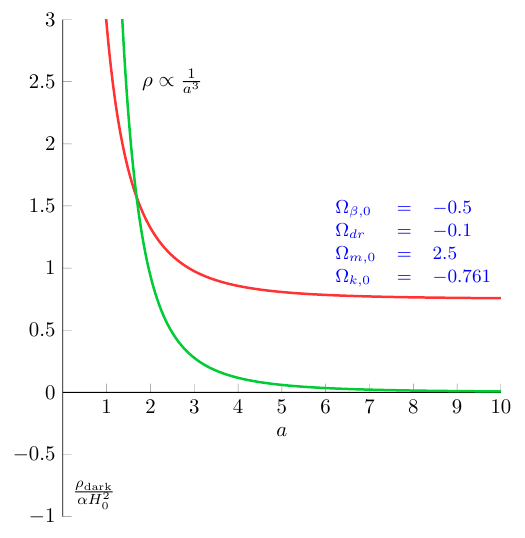}
    \caption{}
    \label{fig:y3}
\end{subfigure}
\caption{The effective $\rd$ contained in the possible
brane closed universe with some parameters
choices. The last plot is a comparison between $\rd$
and the primitive energy $\rho \thicksim 1/a^3$ }
\label{fig2}
\end{figure*}

In Figure~(\ref{fig2}) we depict $\rd$ with $n=0$, for illustration, 
for some chosen values of the $\Omega$s. In both cases, we observe 
that, while $\rho(a)$ explodes as $a \longrightarrow 0$, the 
effective energy density remains finite and positive. Notice
also that, for some specific choice of the energy density 
values, $\rd$ decreases slower than the primitive energy
$\rho \thicksim 1/a^3$. 

It is worth noting that this universe, free of a cosmological 
constant, does not provide relevant information at early times 
making it unrealistic at this stage. To show this it is 
convenient to explore the $\rho=0$ scenario which is dominated 
only by the parameters $\omega$ and $\beta$. By specializing 
the cubic equation~(\ref{rdcubic1}) with $\rho=0$ and solving 
for $\mathcal{Z}$ in addition to focusing on the real solution, 
we are able to determine, after an expansion around $a=0$, that 
the significant contribution to $\rd$ is provided by negative 
values of $\omega$, given by
\be
\n
\rho_{\text{\tiny dark}} \simeq 
\frac{|\omega|^{2/3}}{a^{8/3}}.
\ee
Therefore, the possibility of obtaining a premature model is not 
possible since this result, which agrees with the one 
discussed in~\cite{Davidson2001} under similar conditions, 
this type of universe evolves similarly as a matter dominated 
FRW universe which does not represent a realistic universe. In 
this sense, it seems that the model only has an impact for late 
times since it reproduces an accelerated behaviour.
 We remark in passing that, the non-standard Friedmann equation
in our approach is provided by $\dot{a}^2 + k = f(a;\Ok, \Om, 
\Ob, \Odr)a^2$ where the energy density parameters are given
in~(\ref{denergies}), and we have additionally introduced the energy
density parameter $\Ok := - \frac{k}{H_0^2}$,~\cite{Rojas2012}. 
One can go one step further and cast this Friedmann-like equation 
in the suggestive form $\dot{a}^2 + U(a;\Omega_s) = 0$ where 
$U$ stands for an effective potential parameterized by the density 
energies $\Omega_s$,~(\ref{denergies}),~\cite{Rojas2012}. This 
fashion is convenient since it allows the system to be faced as 
an equivalent one-dimensional non-relativistic mechanical problem 
in which the total mechanical energy is zero. Certainly, we have

\begin{figure*}[!hbt]
\begin{subfigure}{0.41\linewidth}
    \centering
    \includegraphics[width=5.0cm,height=5.5cm,angle=0
    ]{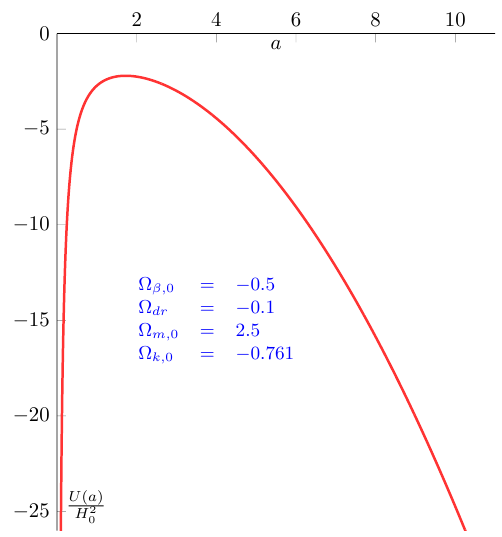}
    \caption{}
    \label{fig:x1}
\end{subfigure}
\quad 
\begin{subfigure}{0.41\linewidth}
    \centering
    \includegraphics[width=5.0cm,height=5.5cm,angle=0
    ]{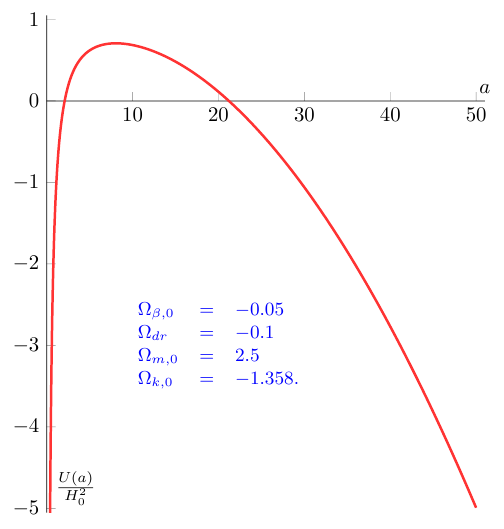}
    \caption{}
    \label{fig:x2}
\end{subfigure}
\caption{The effective potential $U$ describing possible
brane trajectories for a closed universe and some parameters
choices.}
\label{fig1}
\end{figure*}

\begin{figure*}[!hbtp]
\begin{subfigure}{0.42\linewidth}
    \centering
    \includegraphics[width=5.5cm,height=5.5cm,angle=0
    ]{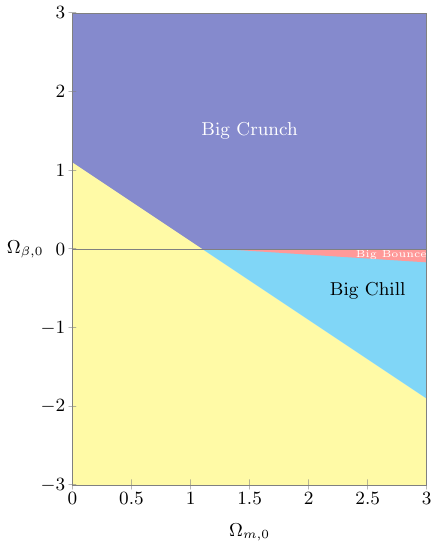}
    \caption{}
    \label{fig:xx1}
\end{subfigure}
\caption{We depict several possibilities for the expansion
of this closed universe in dependence of $\Ob$ and $\Om$. The 
bluish-purple color region is associated with a Big Crunch 
evolution of the universe, and the coral color region corresponds 
to a Big Bounce evolution whereas the blue area is in accordance with
a universe with a Big Chill behavior. The yellow region
corresponds to non-physical universes.}
\label{fig2a}
\end{figure*}

\begin{widetext}
\be 
\frac{U}{H_0^2} =  - \Ok  
- \frac{a^2}{9} 
\left\lbrace 2 \left(\Ob^2  + \frac{3\Om}{a^3}\right)^{1/2}  
F \left[ 
\frac{1}{3} F^{-1} \left( - \frac{\Ob \left( \Ob^2 + \frac{9}{2}
\frac{\Om}{a^3}\right) + \frac{27}{2} \frac{\Odr}{a^4}}{\left( 
\Ob^2 + \frac{3\Om}{a^3} \right)^{3/2}} \right) \right]
- \Ob \right\rbrace^2,
\ee
\end{widetext}
where $F = \cosh x$ for $|x| > 1$, or $F= \cos x$ for $|x| 
\leq 1$. In passing, regarding the values of the energy 
densities~(\ref{denergies}), we emphasize that these are 
restricted to obey the normalization condition
\be 
\label{normalization}
\left( 1 - \Ok \right)^{1/2} \left( \Ok + \Om - 1\right)
- \Ob \left( 1 - \Ok \right) = \Odr.
\ee
This formula will prove very useful to carry out numerical
analysis as well as analyze special limits. Indeed, for 
$\Odr = \Ob = 0$ together we immediately recover the usual 
FRW cosmology whereas for only $\Odr = 0$, we recover the DGP 
approach admitting the two well-known branches of the 
theory,~\cite{DGP2000b,DGP2000a,Rojas2012}. 
We depict this potential for some chosen values of the energy
densities in Figure~(\ref{fig1}). In both cases we observe
two general features: the singularity of $U(a,\Omega_s)$
at small scales, $a \to 0$, as well as at large scales
$a\to \infty$. More precisely, in dependence on the chosen
parameters we find either a universe expanding forever 
(Big Chill), which is a signal of the presence of mysterious 
dark energy in order to counteract the gravitational 
interaction or else the universe may present a Big Bounce 
destiny, that is, in a state of constant expansion 
(Big Bang) and eventually reverses and re-collapses 
(Big Crunch). In Figure~(\ref{fig2a}) we have a graph
enclosing different possibilities for the universe expansion
in our approach depending upon the densities $\Ob$
and $\Odr$.

\begin{figure*}[!hbt]
\begin{subfigure}{0.4\linewidth}
    \centering
    \includegraphics[width=\linewidth]{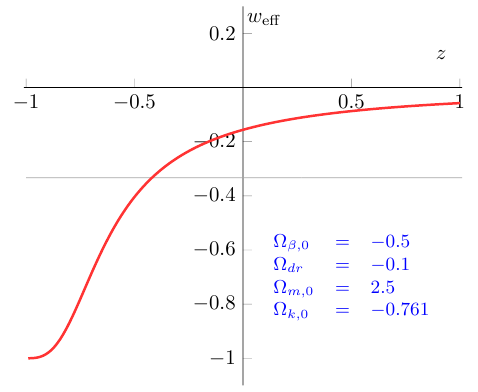}
    \caption{}
    \label{fig:z1}
\end{subfigure}
\qquad \quad
\begin{subfigure}{0.4\linewidth}
    \centering
    \includegraphics[width=\linewidth]{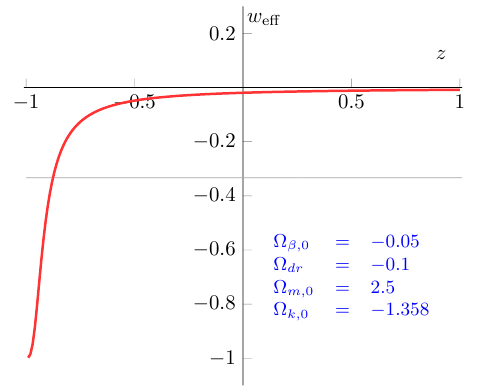}
    \caption{}
    \label{fig:z2}
\end{subfigure}
\caption{The state parameter is depicted for some values. 
The graph on the left corresponds to a Big Chill behavior,
describing a self-acceleration universe. On the contrary,
the right plot corresponds to a Big Bounce dynamics.}
\label{fig3}
\end{figure*}

Having disclosed the unconventional behavior of the dark 
energy content in our approach, in accordance with our 
symmetry considerations we can assume the resulting 
effective fluid in~(\ref{FRW1}) to have an equation 
of state of the form $\Pe = \ome \re$, where $\ome$ 
is subject to the values of the $\Omega$s. Indeed,
by assuming $\Pe = - \left[ \frac{a}{3\dot{a}} \left( 
\dot{\rho} + \dot{\rd} \right) + \left( \rho +
\rd \right) \right] / \left( \rho + \rd \right)$, 
using~(\ref{Pd}),~(\ref{sol4}) with $n=0$ for simplicity, 
as well as the redshift $z$ expression when the universe 
was $1/(1+z)$ of its present size, it follows that
\begin{widetext}
\be
\begin{aligned}
\ome = - 1 +  \frac{3(1+z)^3}{\oall^2 \Pall} 
& \left\lbrace   
\Om (\Ob + \Pall) \right.
\\
& \left. + \frac{3 (1+z)\oall \left[ (1+z)^2\Om^2 \Ob 
+ \Odr [(1+z)^3 \Om - \frac{8}{3}\Ob^2] \right]\sin 
(\varphi_{m,\beta,dr,z})}{\sqrt{\oall^6 - \Oall^2}}
\right\rbrace,
\end{aligned}
\ee
\end{widetext}
where
\be
\begin{aligned} 
\Oall &:= - \Ob \left[ \Ob^2 + \frac{9}{2} (1+z)^3 \Om 
\right] 
\\
& - \frac{27}{2} (1+z)^4 \Odr,
\\
\oall &:= \left[  \Ob^2 + 3 (1+z)^3 \Om \right]^{1/2},
\\
\varphi_{m,\beta,dr,z} &:= \frac{1}{3} \cos^{-1} 
\left( \frac{\Oall}{\oall} \right),
\\
\Pall &:= - \Ob + 2 \oall \cos \left( 
\varphi_{m,\beta,dr,z}
\right).
\end{aligned}
\label{defs}
\ee
It is worthwhile to note that our results specialize to
the GR limit for $\Odr = \Ob = 0$. Indeed, for such a case
we find $\Omega_{m,0,0,z} = 0, \omega_{m,0,z} = 
\sqrt{3(1+z)^2} \,\Om, \varphi_{m,0,0,z} = \pi/6$ and 
$\Phi_{m,0,0,z} = \sqrt{3}\,\omega_{m,0,z}$, so that, 
with these values we obtain $\ome = 0$ as in the standard 
cosmology. One can show that for $\beta = 0$, at least numerically, 
the state parameter $\left.\ome \right|_{\beta= 0}
> -\frac{1}{3}$ for allowed physical regions, that is, 
we do not have acceleration behavior. In some sense, 
this explains the reason why the non-vanishing values of 
$\Odr$ and $\Ob$ induce self-(non-self)-acceleration for 
this type of universe at late times.
In Figure~(\ref{fig3}) we plot the state parameter for 
some chosen values of the energy densities $\Omega_s$. 
In comparison with the standard cosmology, we observe the 
presence of dark energy content because $\ome <0$.

We further observe that for the chosen values, and at the present 
time, we do not have a complete energy content to 
fulfill the condition $-1 < \ome < -1/3$. It is mandatory to 
choose either another set of energy density values or the 
inclusion of some matter fields, in order to get a 
self-acceleration behavior to reproduce the known observations
for the dark energy. Additionally, it is worthwhile to observe
that at late times the combination of the energy density values
mimics, in some sense, the role of a cosmological constant. This is 
so because we are able the reach the value $\ome = -1$ in our
plot, which is a characteristic of a de Sitter type universe. Further, 
the state parameter never crosses the effective
phantom line, $\omega_{\text{\tiny eff}}$. As a consequence, 
with the chosen values $\Omega_s$, the usual dominant energy 
condition is not violated in agreement with the non-negative 
property of $\re$.

\begin{figure*}[!hbtb]
\begin{subfigure}{0.41\linewidth}
    \centering
    \includegraphics[width=5.0cm,height=5.5cm,angle=0
    ]{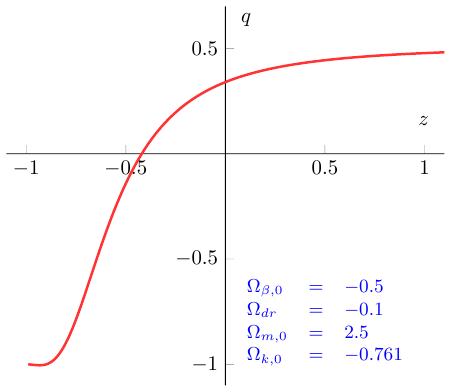}
    \caption{}
    \label{fig:w1}
\end{subfigure}
\qquad \quad
\begin{subfigure}{0.41\linewidth}
    \centering
    \includegraphics[width=5.0cm,height=5.5cm,angle=0
    ]{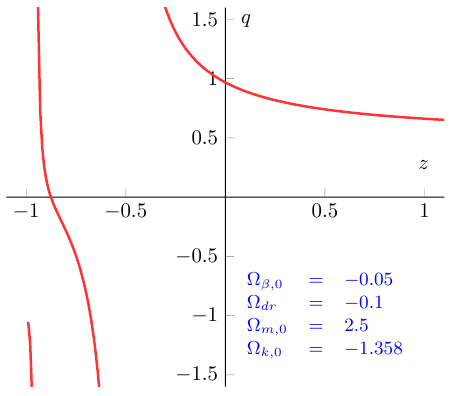}
    \caption{}
    \label{fig:w2}
\end{subfigure}
\caption{The deceleration parameter is depicted for the values
$\Ob =-0.5$, $\Odr = - 0.1$, $\Om = 2.5$ and $\Ok = -0.761$. 
For the left plot, the transition value is given when 
$z\approx -0.419$, which corresponds to the value 
$a \approx 1.72$ (See figure~(\ref{fig1}) for comparison.).}
\label{fig4}
\end{figure*}

Another parameter worth exploring is the deceleration parameter, 
$q(a) = - 1 - \dot{H}/H$.  
After a lengthy but straightforward computation, we find that
\begin{widetext}
\be
\begin{aligned} 
q = -1 + \frac{9 (1+z)^2}{\Pall^2 + 9(1+z)^2 \Ok} 
& \left\lbrace 
\Ok + \frac{(1+z)}{2} \frac{\Pall}{\oall^2} \left[
\Om (\Ob + \Pall)
\right. \right.
\\
& - \left.\left. 
\frac{2\oall\left[ \left( 4(1+z) \Odr  + \Om \Ob \right) \oall^2 
+ \Om \Oall \right]\sin (\varphi_{m,\beta,dr,z})}{\sqrt{\oall^6 
- \Oall^2}}
\right]
\right\rbrace,
\end{aligned}
\ee
\end{widetext}
where we have considered the structures~(\ref{defs}). We observe
immediately that the transition value when the deceleration turned
into acceleration is provided by $\Pall = 0$. Additionally,
the GR limit is reached for $\Odr = \Ob = 0$, that is,
$q = \frac{\Om/a^3}{2(\Om/a^3 + \Ok/a^2)}$, as usual. In 
figure~(\ref{fig4}) we depict the $q$ deceleration parameter 
for some chosen values of the $\Omega_s$. There, for a Big Chill
expansion we found a typical $q$ deceleration behavior where
the transition redshift $z_0$, for some given values for the
$\Omega_s$, is $z_0 \approx - 0.419$. On the contrary, for a 
Big Bounce expansion we observe that the interval $-0.52 < z < 0.95$
does not allow a physical scenario which is in agreement with the
interval $2.09 < a < 21.214$ found for the corresponding
potential plot in Figure~(\ref{fig1}).

\subsection{DGP-like dark energy from modified RT cosmology}

\begin{figure*}[!hbt]
\begin{subfigure}{0.41\linewidth}
    \centering
    \includegraphics[width=5.0cm,height=5.5cm,angle=0
    ]{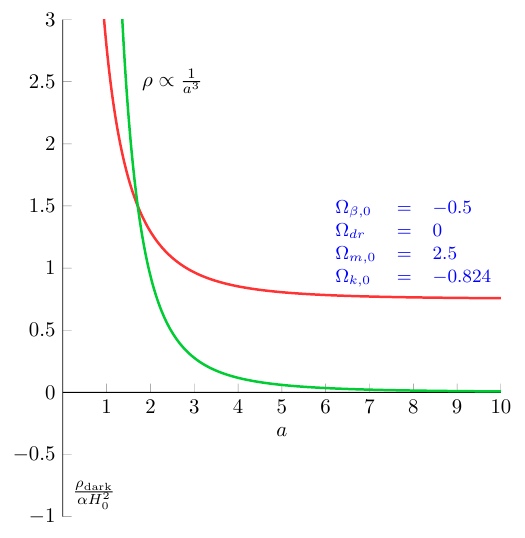}
    \caption{}
    \label{fig:s1}
\end{subfigure}
\qquad
\begin{subfigure}{0.41\linewidth}
    \centering
    \includegraphics[width=5.0cm,height=5.5cm,angle=0
    ]{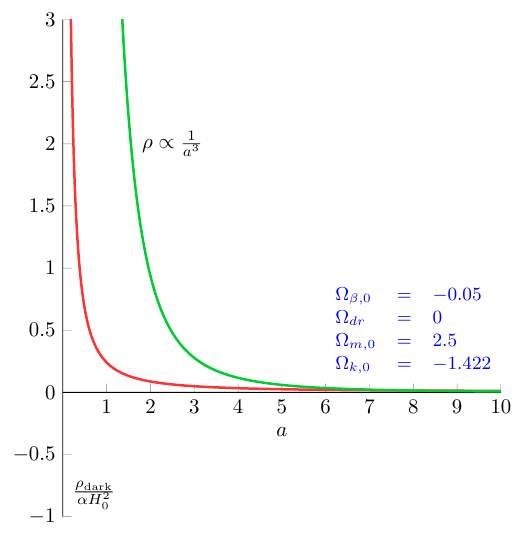}
    \caption{}
    \label{fig:s2}
\end{subfigure}
\caption{The dark energy $\rd$ for the DGP approach 
considering some values of the energy densities.
In both cases, we compare the behavior of $\rd$ with 
the primitive energy density $\rho \sim 1/a^3$.}
\label{fig6}
\end{figure*}

The radiation-like energy density, $\Odr$, represents the 
fingerprint of the extra-dimensional ambient spacetime. As 
we already discussed, it influences the dynamics of this 
brane-like universe. At the limit
$\Odr = 0$, the master equation~(\ref{eom4}) reduces remarkably
to $3\alpha (\dot{a}^2 + k) + 3\beta \,a (\dot{a}^2 + k)^{1/2}
- \rho\,a^2 = 0$. When inserting~(\ref{FRW1}) and recalling
that $\re = \rho + \rd$, we have that $\re = \rho - \beta_* 
\sqrt{\rho + \rd}$. We determine immediately that
\be 
\rd= -3 \beta \sqrt{H^2 + \frac{k}{a^2}}.
\label{rhoDGP}
\ee
This special limit assures that the effective dark energy
within this geodetic brane context is nothing more than 
a geometric effect provided by the correction term $K$. Obviously, we 
can include a cosmological constant in our development
but we claim to highlight the role of the inclusion of the
trace $K$ model. Additionally, from this expression, we immediately 
identify that $\beta$ parameter plays the role of the inverse
of the crossover scale of the DGP 
approach~\cite{DGP2000b,Maartens2010}. Nonetheless, a word 
of caution is needed; although $\beta$ can take positive and
negative values, it is mandatory that $\rd$ be subject to
the condition $\re \geq 0$. In Figure~(\ref{fig6})
we depict $\rd$ for this special case. 

\section{Concluding remarks}
\label{sec4}

In this paper, we have analyzed the dark energy content 
within the geodetic brane gravity context enhanced with 
an extrinsic curvature term associated with the shape of 
the brane. We provide evidence that such dark energy can 
originate from the correction of extrinsic curvature terms 
appearing in the action functional that models our Universe. 
To show this we have considered a FRW geometry defined 
onto our braneworld probing a $5$-dimensional Minkowski 
background spacetime. Our approach is free of a cosmological 
constant with the idea of highlighting the role that the 
correction term $K$ plays in development. In this sense, it 
plays the role of a cosmological constant providing self-(and 
non-self) acceleration behavior at late times. One can think of the $K$ 
term as a relic or a possible fingerprint of the existence 
of a $5$-dimensional EH action that disappears 
at some time. Regarding this point, on geometrical grounds, it is worth noting that including 
the $K$-term is only possible for timelike oriented hypersurfaces floating in a background spacetime. Indeed, we can not lift this restriction and consider an arbitrary co-dimension because it would be impossible to maintain the symmetry of the action functional. By assuming a traditional point of view for the 
Friedmann equation, we observed that the regular matter is 
accompanied by an atypical piece, termed $\rd$. In our 
development, the intriguing dark energy effect results 
as a geometrical achievement originated by the model under 
consideration. 
What is notable is to observe how the correction terms 
associated with the extrinsic curvature of the brane can 
reproduce many of the general characteristics of the so-called 
dark energy.

Our approach is not intended to be realistic, but on the 
contrary, it aims to be the promoter of extrinsic curvature 
corrections that help to understand several physical 
phenomena ranging from condensed soft matter to cosmology 
and particle physics. Because of this, perhaps the inclusion 
of some matter fields, such as the cosmological constant, 
can take on a more realistic meaning. It would be interesting to extend the 
analysis of our cosmological findings and verify to what
extent the results obtained are close to current observational data such as those provided by the $\Lambda$CDM model. This involves, for example, graphing the different $\Omega$s as functions of Log$[a]$ for comparison, but this would involve modifying some starting expressions. This is however a more deep numerical analysis that is beyond the scope of this note. A consequent extension 
of our work presented here is the incorporation of other 
attractive extrinsic curvature terms such as those provided 
by the so-called Lovelock brane gravity theory. For instance, 
the Gibbons-Hawking-York-Mayers type term, $K^3 -2 K 
K_{ab}K^{ab} + 3 K^a{}_b K^b{}_c K^c{}_a$, \cite{Rojas2013}. 
We believe that such an addition could improve or enrich 
the cosmological findings in this braneworld context.  This 
will be reported elsewhere.

\begin{acknowledgments}
We would like to thank anonymous 
referees for constructive comments and for 
drawing our attention to some missing references. The authors are grateful to R. Cordero, J. A. M\'endez Zavaleta
and M. Cruz for discussions and helpful suggestions. 
ER acknowledges encouragement from ProDeP-M\'exico, 
CA-UV-320: \'Algebra, Geometr\'\i a y Gravitaci\'on.
JCNZ acknowledges support from a CONAHCYT-M\'exico 
graduate fellowship. GC acknowledges support from
a Postdoctoral Fellowship by Estancias Posdoctorales por M\'exico 
2023(1)-CONAHCYT . Also, ER thanks partial support 
from Sistema Nacional de Investigadores, M\'exico.
\end{acknowledgments}

\appendix

\section{Extended objects geometry}
\label{app1}

Consider a $(p+1)$-dimensional manifold $m$ that represents 
the worldvolume, $m$, of a spacelike brane $\Sigma$. $m$ is 
embedded in a $N$-dimensional flat Minkowski spacetime, $\mathcal{M}$, 
with metric $\eta_{\mu\nu}
=\text{diag}(-1,1,...,1)$ ($\mu,\nu = 0,1,2, 
\ldots,N-1$). The worldvolume is described by the embedding 
functions $X^{\mu}(x^a)$ where $x^a$ are  local coordinates 
for $m$ ($a,b = 0,1,2, \ldots p$). The tangent vectors to 
$m$ are given by $e^\mu{}_a: = \partial X^\mu/\partial x^a$
which allow us to define an induced metric on $m$ as 
$g_{ab}:=\eta_{\mu\nu} e^{\mu}{}_a e^{\nu}{}_b$. 
As usual, $g^{ab}$ denotes the inverse of $g_{ab}$, 
and $g = \det (g_{ab})$. Additionally, $m$ is assumed to be 
time-like so $g < 0$. The $i$th normal vector to $m$, 
$n^{\mu\,i}$, ($i,j=1,2,...,N-p-1$), is defined by the relations 
$n^i\cdot e_a=0$ and 
$n_i \cdot n_j=\delta_{ij}$. These expressions define the normal 
vectors up to a sign and a local $O(N-p-1)$ rotation. This 
gauge freedom requires a gauge field, known as \textit{twist 
potential}, given by $\omega_a{}^{ij} = - n^i \cdot 
\partial_a n^j$,~\cite{defo1995}. In the case of a 
hypersurface embedding, $i=1$ and the extrinsic twist vanishes 
identically. Further, $\nabla_a$ denotes the (torsionless) covariant
derivative compatible with $g_{ab}$. 

For codimension one case, ($i=1$), the extrinsic curvature 
of $m$ is  $K_{ab} = 
- n \cdot \nabla_a \nabla_b X$, and the mean extrinsic 
curvature as its trace, $K = g^{ab} K_{ab}$.  
The corresponding Gauss-Weingarten equations are
\be 
\begin{aligned}
D_a e^\mu{}_b &= \Gamma^c_{ab} e^\mu{}_c - K_{ab} n^\mu,
\\
D_a n^{\mu} & = K_{ab} g^{bc}e^\mu{}_c ,
\end{aligned}
\ee
where $D_a = e^\mu{}_a D_\mu$ with $D_\mu$ is the covariant derivative
compatible with a possible background metric, say $\mathcal{G}_{\mu\nu}$, and
$\Gamma^a_{bc}$ are the connection coefficients compatible
with $g_{ab}$. The intrinsic and extrinsic geometries of the worldvolume 
$m$ are related by some integrability conditions. In 
this sense, the Gauss-Codazzi, and the Codazzi-Mainardi 
integrability conditions are
\beq 
0 &=& \R_{abcd} - K_{ac}K_{bd} + K_{ad}K_{bc},
\label{GCequation}
\\
0 &=& \nabla_a K_{bc} -\nabla_b K_{ac}.
\label{CMequation}
\eeq





\end{document}